\documentstyle[12pt,epsfig]{article}
\def\ltsim{\raise0.3ex\hbox{$<$\kern-0.75em\raise-1.1ex\hbox{$\sim$}}}
\def\gtsim{\raise0.3ex\hbox{$>$\kern-0.75em\raise-1.1ex\hbox{$\sim$}}}

\long\def\comment#1{ }    

\begin{document}
\begin{titlepage}
\begin{flushright}
NUC-MINN-01/16-T\\
LBNL-47837\\
\end{flushright}
\begin{centering}
\vfill

{\bf
Photon emission in heavy ion collisions at the CERN SPS
}

\vspace{0.5cm}
 P.~Huovinen$^{\rm a,b,}$\footnote{huovinen@physics.umn.edu}
 P.V.~Ruuskanen$^{\rm c,}$\footnote{vesa.ruuskanen,
 sami.rasanen@phys.jyu.fi}
 S.S.~R\"as\"anen$^{\rm c,2}$

\vspace{1cm}
{\em $^{\rm a}$Lawrence Berkeley National Laboratory, Berkeley, CA
94720, USA}\\
\vspace{0.3cm}
{\em $^{\rm b}$School of Physics and Astronomy, University of Minnesota,\\
               Minneapolis, MN 55455, USA}\\
\vspace{0.3cm}
{\em $^{\rm c}$Department of Physics, University of Jyv\"askyl\"a,
P.O.Box 35, FIN-40351
Jyv\"askyl\"a, Finland}\\

\vspace{1cm}
\centerline{\bf Abstract}
\end{centering}

We compute the thermal photon spectrum in the Pb+Pb collisions at the
CERN SPS energy using thermal emission rates and a hydrodynamic
description for the evolution of produced hot matter and compare our
results with the measurements of the excess photons by the WA98
collaboration.  Our results show that the measured photon spectrum can
be reproduced with realistic initial conditions which take properly into
account also the finite longitudinal size of the initial collision zone
and which simultaneously describe well both the transverse and
longitudinal hadron spectra.  In the scenario with initial formation of
QGP the recently calculated emission rate, complete to order $\alpha_s$,
reproduces the measured spectrum.  However, the experimental spectrum
can also be reproduced in a purely hadronic scenario without transition
to QGP state, but a high initial temperature, much over the values
predicted for the phase transition temperature $T_c$, is required.

\vfill
\end{titlepage}

\setcounter{footnote}{0}

\section{Introduction}

For the study of the properties of matter created in a high energy
heavy-ion collision, photons and lepton pairs provide the most direct
probe since, after being emitted, they escape the system without further
interactions.  Unfortunately, electromagnetic interactions occur during
all the different stages of the collision, ranging from those between
the incoming partons to the decays of final hadrons, and it is difficult
to disentangle experimentally the contributions from these different
sources.

After a painstaking analysis, the WA98 collaboration has succeeded in
isolating an excess of photons over those originating from the decays
of final hadrons \cite{WA98a,WA98b}. Since a QCD calculation of photon
production from the primary interactions of incoming partons fails to
reproduce this excess\footnote{ It should be kept in mind that a
 consistent description of inclusive photon data in hadronic
 collisions in fixed target experiments is difficult to achieve
 \cite{Aurenche1999}.}
\cite{WA98b,WW}, they can be associated with the secondary interactions
of final state particles. In a phenomenological analysis of heavy-ion
collisions, based on the assumption of the formation of a thermal
system, these are often called the thermal photons.

Since the publication of the WA98 data there has been several attempts
to describe the data using a thermal description for the expanding
matter~\cite{Srivastava,Alam,Peressounko,Chaudhuri,Gallmeister} and
the calculated thermal emission rates of photons.In these studies the
time evolution of matter has either been given by a simple
parameterization of spherical expansion~\cite{Gallmeister}
or described by using boost invariant
hydrodynamics~\cite{Srivastava,Alam,Peressounko,Chaudhuri}.  In the
papers applying hydrodynamics, the reproduction of the data has
required either large initial temperature which led to unrealistically
short initial time for boost invariant flow~\cite{Srivastava},
significant initial radial velocity~\cite{Alam,Peressounko,Chaudhuri},
or large modifications in hadron masses~\cite{Alam}.

In the papers applying hydrodynamics~\cite{Srivastava,Alam,Peressounko,
Chaudhuri} the validity of boost invariant expansion is taken at face
value.  However, from the hadron data at the SPS, we know that the
rapidity window where the expansion is boost invariant is narrow if it
exists at all.  Even if the flow is approximately boost invariant at
later times, it can be expected to deviate from the scaling behaviour
for $\tau \ltsim 1$ fm/c because of the
finite longitudinal extent of the collision zone as discussed in more
detail below.  It is also known~\cite{Sollfrank:1999} that if the
assumption of boost invariance is relaxed and the longitudinal expansion
is constrained to reproduce the measured rapidity spectra of hadrons,
the initial energy densities become strongly peaked and larger by a
factor two than the Bjorken estimate. As is evident in~\cite{Srivastava}
the thermal photon production at large $k_T$ is very sensitive to initial
temperature and dominated by the earliest moments of the expansion.
Therefore, the photon production even at midrapidity becomes sensitive
to the details of the initial state including its longitudinal structure.

In this letter we take a conservative approach to photon
production. We use well known vacuum properties of mesons also in
dense medium and set the initial radial velocity to zero.  Instead of
studying these effects we concentrate on the effects on photon spectra
caused by different initial states and equations of state (EoS) of the
expanding matter. In all cases the calculation is constrained to
provide a reasonable description of both transverse and longitudinal
hadronic spectra~\cite{Huovinen:1999}, not only the final particle
multiplicity as in~\cite{Srivastava,Alam,Chaudhuri}. In our model
finite baryon number density is included in the EoS. We assume local
chemical equilibrium with zero strangeness density until kinetic
freeze-out. This allows a reasonable reproduction of charged particle
and net proton spectra but fails to describe the details of strange
particle yields. However, this is not a severe problem since
reproduction of negative particle spectra means the description of the
bulk of the matter is realistic. The processes involving strange
particles are not included in the photon production rates in hadron
gas either. To compare with the results in literature, we calculate
the photon spectra also with boost invariant expansion. As thermal
production rates we use rates provided in the literature. For the QGP
we use the recently obtained full order of $\alpha_s$
result~\cite{Arnold:hep-ph/0109064,Arnold:hep-ph/0111107} for zero
quark chemical potential, $\mu_{q}=0$.  This result contains both the
lowest order one-loop rates~\cite{Kapusta,Baier} with exact
non-logarithmic term and completes the earlier two-loop calculations
\cite{Aurenche,Thoma} through summation of leading contributions from
multi-loop terms up to all orders.  In hadron gas we use the
parameterizations~\cite{Nadeau:1992,Brown} of the rates calculated
using effective Lagrangians and vector-dominance
model~\cite{Kapusta,Brown}.

\section{The framework for calculating thermal spectra}

Consider first the emission rates of photons from a thermal matter.
Since we assume that the photons escape the system after emission the
emitted spectrum is not the thermal, black-body radiation spectrum.

In the QGP one is faced with the problem of infrared singularities in
the rate calculation. In lowest order this leads to a logarithmic
energy dependence on $\omega\slash T$ in addition to the usual
exponential (in Boltzmann approximation) $\exp\{-\omega\slash T\}$
dependence \cite{Kapusta,Baier}.  The next-to-leading order 2-loop
result \cite{Aurenche,Thoma} turned out to be of the same order in
$\alpha_S$ and of the same size or larger than the leading 1-loop
result.  It was argued that also the leading parts of the higher-loop
contributions are of order $\alpha_s$ \cite{Aurenche2}.  Recently the
resummation of order $\alpha_s$ contributions from all orders in the
multi-loop expansion was achieved, fully completing the order of
$\alpha_s$ analysis of the photon emission rate in the quark-gluon
plasma~\cite{Arnold:hep-ph/0109064,Arnold:hep-ph/0111107}.
In our calculation we use the parameterization of the rate provided in
\cite{Arnold:hep-ph/0111107}.
It should be remembered that this result is for zero baryochemical potential
whereas in the collisions at SPS energy net baryon density is relatively high.
In the leading order, the effect of
non-zero $\mu_q$ has been considered~\cite{Traxler:1994hy} and,
in good approximation, it was found to be included into the rate by
replacing the multiplicative factor $T^2$ by $T^2+\mu_q^2/\pi^2$.
We have checked that in the plasma phase the ratio of these factors
is typically $1.02<1+\mu_q^2/(\pi T)^2<1.05$ indicating only a few per cent
increase to the calculated plasma contributions.

Another problem in applying the calculated rates in nuclear collisions
is that the assumption of high temperature limit with an ideal QGP may
not be valid.  The lattice QCD calculations of energy density and
pressure indicate that the ideal gas or Stefan--Boltzmann limit is
approached slowly with increasing temperature~\cite{Karsch}.  On the
other hand recent lattice calculations on lepton pair emission rate
\cite{Karsch:hep-lat/0110208} show that, at least in some parts of the
phase space, the lowest order (in $\alpha_s$) perturbative result is
quite close to the full lattice result.  This is encouraging for the
type of phenomenological study as presented here and supports the
interpretation of data in terms of results based on perturbative rates.

In the hadron gas the scattering processes $\pi\pi\to \rho\gamma$ and
$\pi\rho\to \pi\gamma$ are included in the photon emission rate.  In
addition to the contributions from interactions described with a
pseudoscalar-vector Lagrangian~\cite{Kapusta}, the process $\pi\rho\to
\pi\gamma$ gets a large contribution from the $\pi\rho$ interaction with
the $a_1$ axial meson:  $\pi\rho\to a_1\to\pi\gamma$~\cite{Brown}.  In
our calculations we use the parameterizations provided by the authors
in~\cite{Nadeau:1992,Brown} for these rates.  Since in the WA98 analysis
the decay contributions of final state hadrons are subtracted, only
those $\rho\to\pi+\pi+\gamma$ and $\omega\to \pi^0+\gamma$ decays which
occur in the hadron gas before freeze-out are included.  Even though the
decays of final $\pi^0$'s and $\eta$'s dominate the decay photons from
final hadrons, their decays are so much slower than those of $\rho$ and
$\omega$ that they can be neglected during the hadron gas phase.
The effects of baryons on the emission rates are not known and our results
contain rates from mesonic processes only. In the hydrodynamic calculation of
expansion baryons are included.

We study the evolution of the system using three different equations of state
(EoS). Two of the EoSs describe a first order phase transition
from hadron gas to QGP with transition temperatures 165 MeV (EoS\,A) and
200 MeV (EoS\,D) and the third one describes the matter as a hadron gas
also at high temperatures (EoS\,H). The QGP is an ideal gas of quarks
with 2.5 flavours and gluons. The hadron gas contains all known
hadrons with masses up to 2~GeV. A more detailed discussion can be
found in~\cite{Huovinen:1999,Sollfrank:1997}.
The present results from lattice calculations for the transition
temperature are $T_c(n_f=3)=154\pm8,\ T_c(n_f=2)=173\pm8,\ {\rm and}\
T_c(n_f=2)=171\pm4$ MeV, where the first two values are for improved
staggered fermions and the last one for Wilson fermions~\cite{Karsch}.
For this reason we consider the EoS\,A with $T_c=165$ MeV as our
reference EoS.

As described in the introduction we want to study the effects of initial
conditions on the longitudinal expansion and the photon production.  For
that purpose we do the calculation using three different descriptions of
initial state.  For two of them, labeled IS\,1 and IS\,2, the
longitudinal expansion is not boost invariant.  To be able to
distinguish what is the effect of longitudinal expansion, we also carry
out the calculations using the boost invariant expansion scenario
(labeled BI in the following).  When the expansion is not boost
invariant, we assume that the rapidity distribution of energy in the
initial state is Gaussian.  The width of the rapidity distribution
increases with the transverse coordinate $r$ since the nuclei become
more transparent towards the edges.  The normalization is fixed by
requiring the energy per unit transverse area in the initial state,
$e(t_0,r)= \int\!dz\,T^{00}(t_0,r,z)$, to equal the energy per unit
transverse area of incoming nucleons~\cite{Sollfrank:1999}.

To convert this distribution to spatial energy density distribution we
have to specify the $z$ dependence of the initial longitudinal velocity.
This choice is by no means unambiguous and since the choice of the
velocity profile affects the maximum initial temperature and thus the
photon yield, we carry out the calculations using two different
profiles.  In both cases the longitudinal size of the system is
constrained to 3.2 fm and the longitudinal flow velocity at the edge of
the system is similar.  In the case of IS\,1 we assume that the
longitudinal flow rapidity increases linearly with distance $z$ from the
center of the system.  In our second choice IS\,2 the dependence is
nonlinear and the velocity is smaller at small $z$, but becomes
at large $z$ larger than in the linear case IS\,1 (see
ref.~\cite{Huovinen:1999}).  In the case of IS\,1 the energy density is
strongly peaked in the middle whereas IS\,2 leads to an almost flat
distribution. Consequently IS\,1 leads to a larger value of the maximum
initial temperature than IS\,2 (see table~\ref{lampo}).
In the following we use IS\,1 as our reference expansion
scenario.

\begin{table}
\begin{center}
\begin{tabular}{|l||c|c||c|c||c||c||}
\hline
       & \multicolumn{2}{c||}{IS\,1} & \multicolumn{2}{c||}{IS\,2}
                                & BI\,(0.3) & BI\,(1.0) \\
                                                             \hline
   & EoS\,A & EoS\,H & EoS\,A & EoS\,H & EoS\,A & EoS\,A \\
                                                              \hline
$T_{\rm max}$ (MeV)
             &  325   &  275   &  265   & 245    &  364   &  244  \\
$\overline{T} (z=0)$ (MeV)
             &  255   &  234   &  214   & 213    &  301   &  214  \\
\hline
\end{tabular} \vspace*{2ex}
 \caption{Maximum value and average initial temperature
          for different EoS and initial state.
          Average temperature is calculated at $z=0$.}
 \label{lampo}
\end{center}
\end{table}

In the case of boost invariant longitudinal flow, $v_z=z\slash t$, we
assume the transverse distribution of energy density
$\epsilon(\tau_0,r)$ to be proportional to the number of participants
per unit transverse area.  Its normalization is chosen to reproduce the
observed pion $p_T$ distribution in most central
collisions~\cite{Kolb:2000}.  The normalization of $\epsilon$ obviously
depends on the initial time
$\tau_0$ which is the time for all the matter in the given
rapidity range to reach local thermal equilibrium.
In the boost invariant scenario this represents the main
uncertainty of the calculation. It should be observed that the hadron
spectra are not very sensitive to $\tau_0$ as long as $\tau_0$ is much
shorter than the transverse time scale $\sqrt 3 R_A$ and if the initial
densities are chosen to give the same total entropy.  However, since the
photon rates have a strong temperature dependence, especially the
high-$k_T$ part of the thermal photon spectrum changes considerably
with different choices of $\tau_0$.
To illustrate this effect, we carry out the boost invariant calculation
for two different initial times, $\tau_0 = 1.0$ and 0.3 fm/c, labeled
BI\,(1.0) and BI\,(0.3), respectively. In both cases we fix the initial
densities by taking the density profile used in~\cite{Kolb:2000} at
$\tau_0=0.8$ fm/c and scale it to initial times $\tau_0=0.3$ and 1.0
fm/c by requiring that entropy per unit rapidity does not change.

At finite collision energy the longitudinal size of the collision zone
is $\sim 2R_A\slash\gamma_{\rm cm}$ which is also the measure of
the time interval for the primary particle production.
Since in the boost invariant expansion the longitudinal size
of the system is $\sim 2\tau$, the initial time should satisfy
$\tau_0\gtsim R_A\slash c\gamma_{\rm cm}\sim 1$ fm/c for a consistent
use of scaling hydrodynamics at SPS.\footnote{In \cite{EKRT}
  a saturation criterion for minijet production is presented leading to a
  time scale $\sim 0.3$ fm/c at the SPS. This is the time scale for the
  production of a minijet of transverse momentum $\sim 0.7$ GeV and, at
  the SPS, cannot be taken as the time interval for the formation of all
  minijets. Indeed, it is not clear if the saturation applies or if soft
  processes with a longer time scale are more significant at the SPS.}
Even though we do not consider initial times as short as
$\tau=0.3$ fm to be realistic for a boost invariant expansion at the SPS,
they may be useful in estimating the limits of photon spectra.
In scenarios without boost invariance, the essential initial parameter
is the longitudinal size of the system.  In fixing the longitudinal size
from the longitudinal collision geometry we have implicitly assumed that
local thermalization times are shorter than this longitudinal scale.

\section{Results}

\begin{figure}[t]
\begin{minipage}[t]{75mm}
  \begin{center}
    \epsfxsize 70mm  \epsfbox{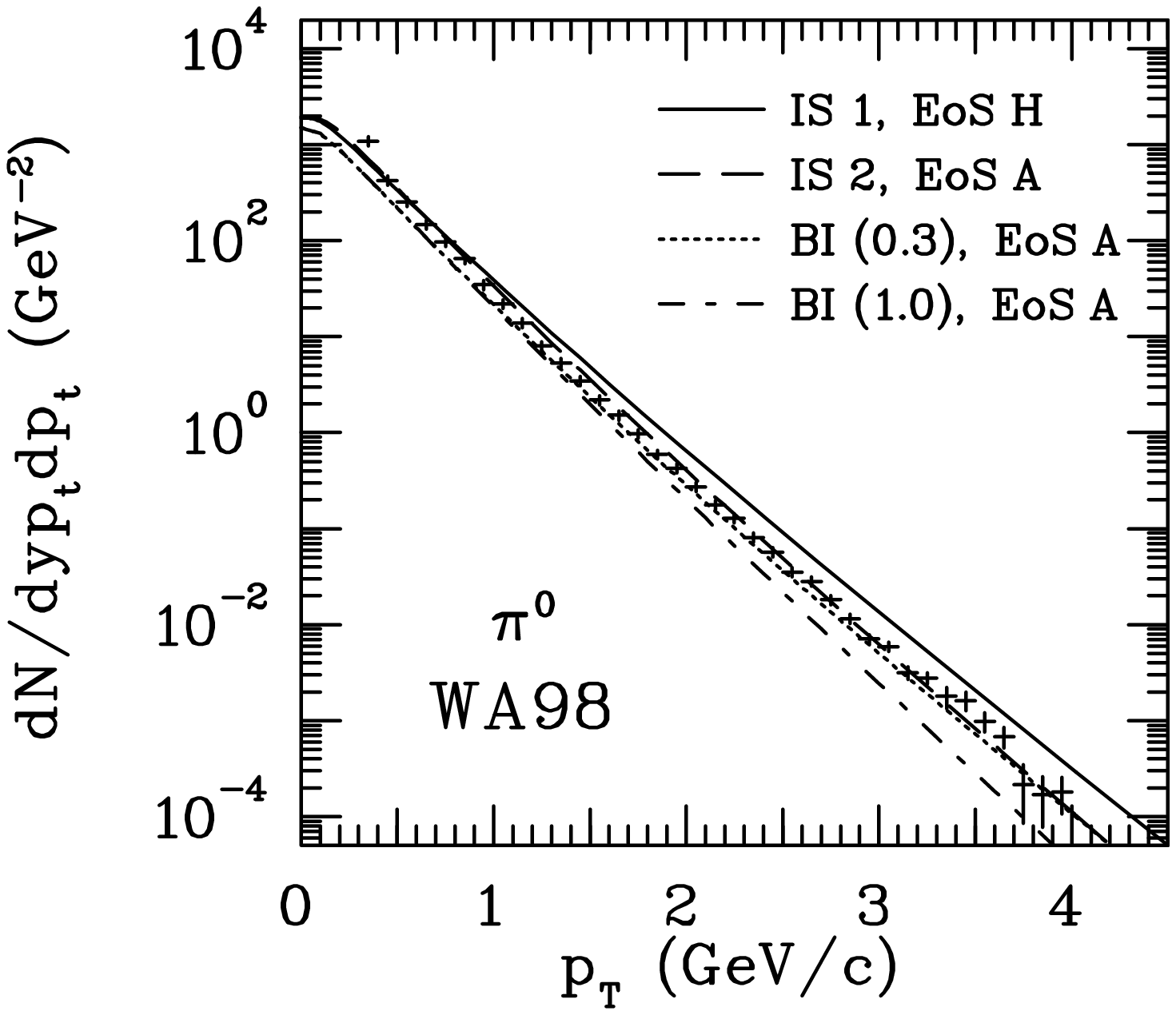}
  \end{center}
 \caption{Calculated $\pi^0 \; p_t$-distribution for different initial states
          and EoSs compared with the
           measurement by the WA98 collaboration~\protect\cite{WA98a}.
           For details, see the text.}
 \label{pi0pt}
\end{minipage}
\hspace{\fill}
\begin{minipage}[t]{75mm}
  \begin{center}
    \epsfxsize 70mm \epsfbox{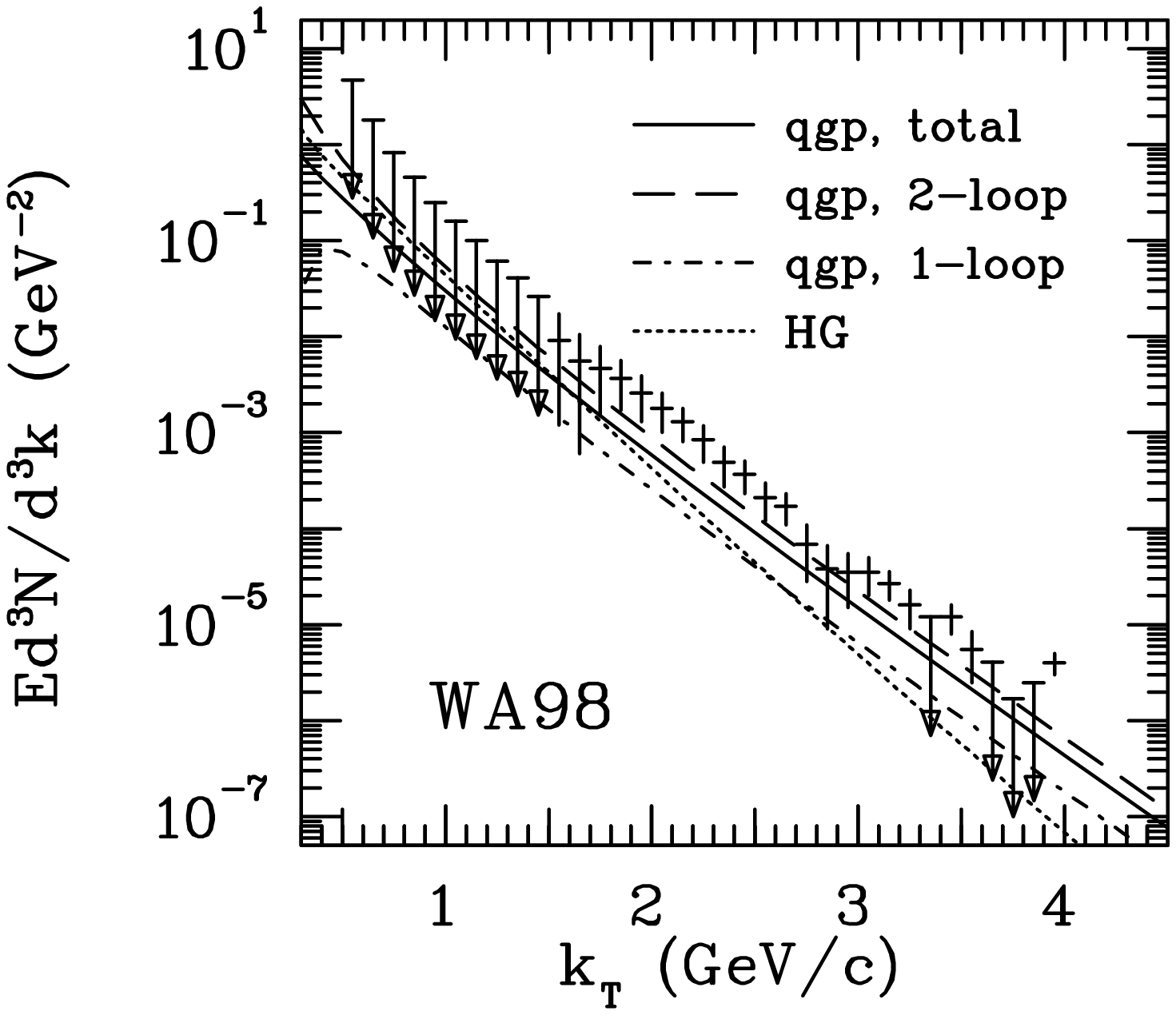}
  \end{center}
 \caption{Thermal photon emission (IS\,1 and EoS\,A) compared with the
      WA98 data~\protect\cite{WA98a}.
      The contribution from plasma is shown for rates calculated in
      leading order (1-loop), next to leading order (2-loop) and in order
      $\alpha_s$ with contributions from multi-loop calculations
      in all orders (total).
      Photons from hadronic phase are shown separately (HG).}
 \label{loops}
\end{minipage}
\end{figure}

We have calculated the photon emission for all the combinations of three
equations of state, EoS\,A, D and H, and the initial conditions,
IS\,1 and IS\,2. For comparison we have performed the calculation also
for the initial conditions BI(0.3) and BI(1.0) with EoS\,A.
To illustrate the quality of our hydrodynamic
description of the nuclear collision, we show in fig.~\ref{pi0pt} the
measured $\pi^0$ spectrum with our calculated results for some of the
cases we have studied. The solid line (IS\,1, EoS\,H) and
the dashed line (IS\,2, EoS\,A) are the two extreme cases for the
calculations with the non-scaling longitudinal expansion. All other
combinations of non-scaling flow and EoS give spectra within these limits.
The dotted (BI\,(0.3), EoS\,A) and dashed-dotted (BI\,(1.0), EoS\,A) lines
represent calculations with boost invariant longitudinal flow, EoS\,A and
initial conditions from \cite{Kolb:2000} scaled to $\tau_0=0.3$ and 1.0 fm,
respectively. We have not tried to fit the $\pi^0$ spectrum but, instead,
used the same initial conditions as in earlier studies of hadron spectra
from several SPS lead-lead experiments~\cite{Huovinen:1999,Kolb:2000}.

In fig.~2 we show the contribution from the hadron gas (dotted line) and
contributions from plasma using the 1-loop rate (dashed-dotted line),
the sum of 1-loop and the order of $\alpha_s$ part of 2-loop rates
(dashed line) and the total order of $\alpha_s$ rate obtained summing
the contributions from multi-loop calculations up to all orders.  The
initial conditions IS\,1 is used in the calculation.  We conclude that
the emission from plasma is dominant for the large transverse momentum
range, $k_T>2$ GeV.  The suppression of the 2-loop result due to the
resummation of higher loop contributions is roughly 30 \% and the 1-loop
contribution is typically $\sim 40$ \% of the total QGP rate.  The
shapes of the three plasma curves are remarkably similar.

In fig.~3 the total photon spectra are depicted for different initial
conditions using EoS\,A in each case.  As discussed in chapter 2, the
initial values of energy density or equivalently of temperature are
smaller for initial conditions IS\,2 than for IS\,1.  In case of IS\,2
the photon emission is reduced as compared to IS\,1, especially at
larger values of $k_T$.  At later times the evolution is quite similar
in both cases and the contributions from hadron gas do not differ
significantly.  For IS\,2 the high-$k_T$ end of the spectrum is not
reproduced.

We show in fig.~3 also the results from calculations with longitudinal
boost invariance starting the evolution either at $\tau_0=1$ or $0.3$
fm/c with the same entropy per unit rapidity, $dS\slash dy$, in both
cases.  With this choice of initial times the results follow closely the
nonscaling calculations so that for $\tau_0=0.3$ fm/c (BI\,(0.3)) the
spectrum is close to that from initial conditions IS\,1 and for
$\tau_0=1$ fm/c (BI\,(1.0)) to that from IS\,2.  The high--$k_T$ data is
reproduced only with high temperature initial states, BI\,(0.3) for the
longitudinal scaling flow and IS\,1 for the non-scaling case.  Since the
cooling due to the longitudinal scaling expansion is very fast in the
case of short initial time, a much higher initial temperature is needed
in case of BI\,(0.3) than in case of IS\,1, see table~\ref{lampo}.

As was argued in the previous chapter, we do not consider the use of a short
initial time, such as $\tau_0=0.3$ fm/c, to provide a fully consistent
description of the early stage of a heavy-ion collision at the SPS
energy even though the flow at mid rapidities at later times might
evolve close to scaling flow. We have seen, however, that the early hot
stage is essential for photon emission emphasizing the need for its
reliable description. One could argue that the longitudinal
size of the collision region is reduced below $\sim 2$ fm by shock
compression but this would also slow down the initial flow from the
scaling velocity $v_z=z/t$.  In fact, this would lead to an initial
state with even higher temperatures and slower expansion velocity than
the one described by IS\,1 and consequently to results overshooting the
measured photon spectrum.

\begin{figure}[t]
\begin{minipage}[t]{75mm}
  \begin{center}
    \epsfxsize 70mm  \epsfbox{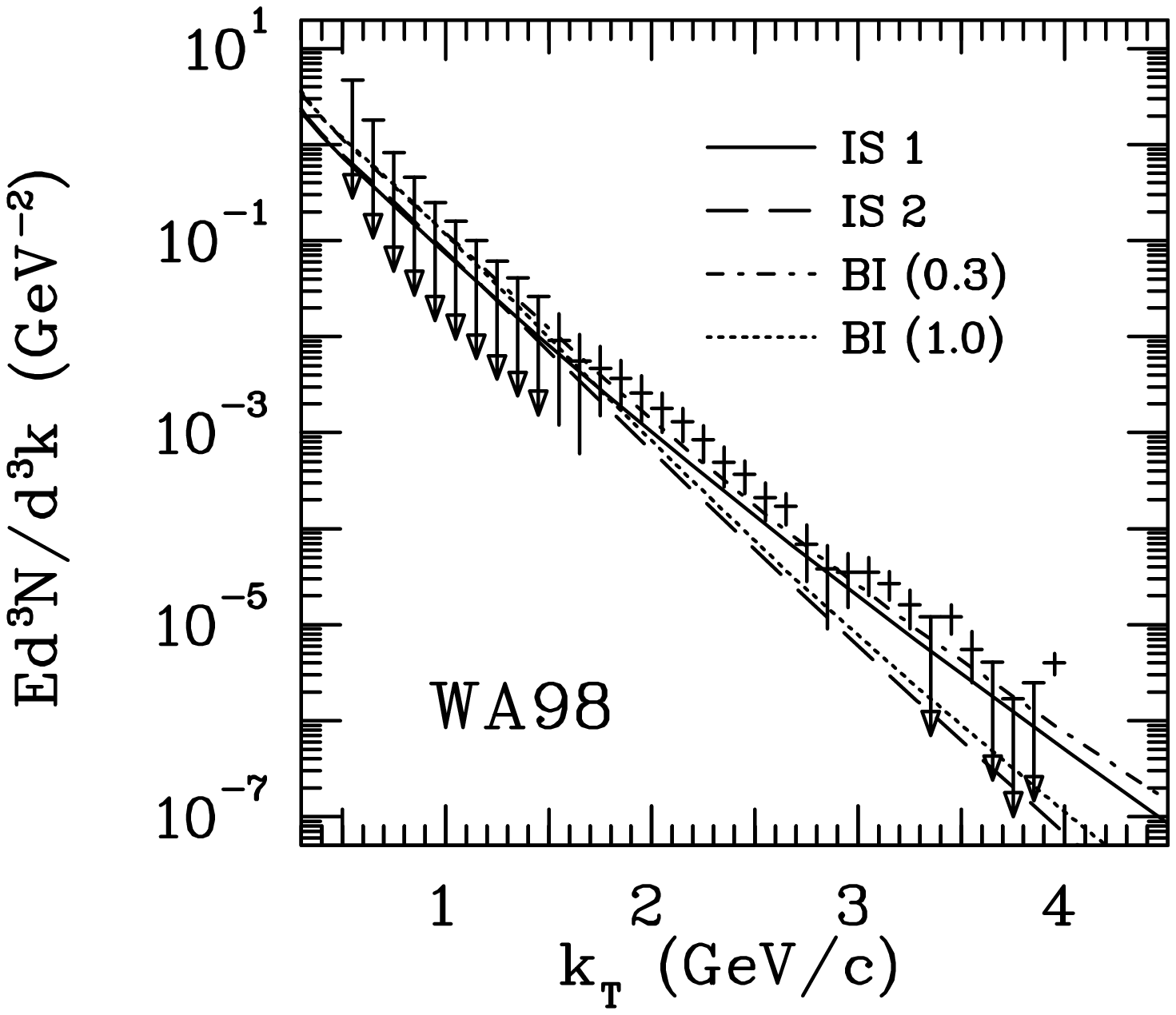}
  \end{center}
 \caption{Thermal photon emission for different initial states.
          Data are from~\protect\cite{WA98a} and calculations are
          done using EoS\,A and total order of $\alpha_s$ rates in the plasma.}
 \label{initial}
\end{minipage}
\hspace{\fill}
\begin{minipage}[t]{75mm}
  \begin{center}
    \epsfxsize 70mm \epsfbox{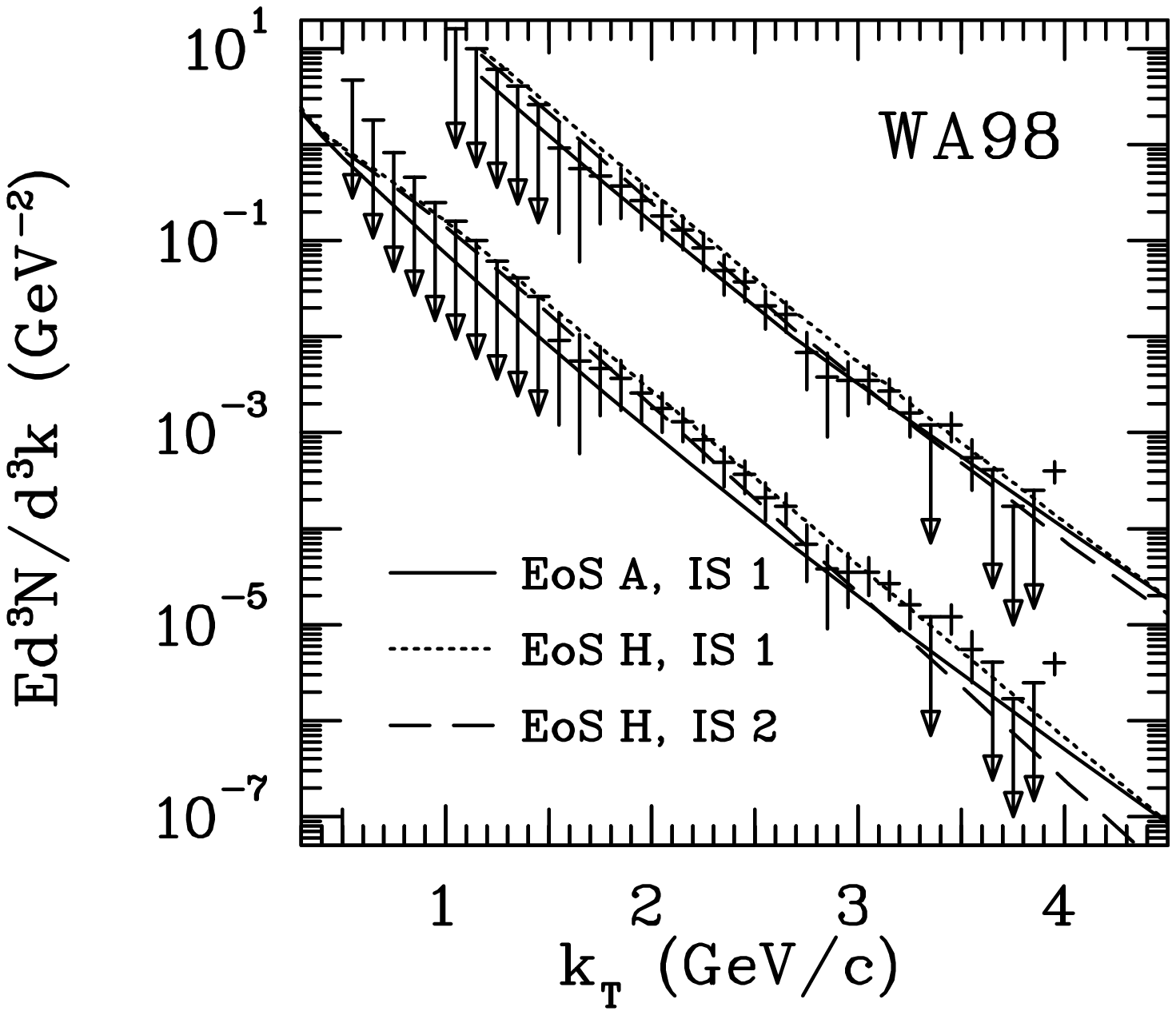}
  \end{center}
 \caption{Thermal photon emission for different EoS and initial state
          with the contribution of initial pQCD photons (upper set)
          and without (lower set). The data and results of the upper
          set have been scaled by a factor 100.}
 \label{eos}
\end{minipage}
\end{figure}

Part of the measured prompt photon spectrum is expected to come from the
primary parton--parton interactions of incoming nuclei.  A perturbative
QCD analysis, which includes also a study of the effect of intrinsic
transverse momentum of partons, gives results which are not more than 30
\% of the measured spectrum when $\langle p_T^2\rangle_{\rm intr} = 0.9$
GeV$^2$~\cite{WW}.  If no intrinsic partonic $p_T$ is included,
calculations are an order of magnitude below the data.  A comprehensive
analysis of prompt photons in hadron-hadron collisions at comparable
energies does not favour large values of intrinsic $\langle
p_T^2\rangle$~\cite{Aurenche1999}.  It looks unlikely that the WA98
photon data can be explained in terms of pQCD photons from interactions
of incoming partons but it would be premature to claim that this
possibility is definitely excluded.

In fig.~\ref{eos} we show the photon spectra for EoS\,A using initial
conditions IS\,1 and for EoS\,H using both initial conditions IS\,1 and
IS\,2.  We have performed the calculations also for the EoS\,D with
$T_c=200$ MeV. Even though the different contributions change, the total
spectra are qualitatively similar for EoS\,A and EoS\,D\footnote{In
        ref.~\cite{Huovinen:1999} the difference between the yields for
        EoS\,A and D was larger than here because we used only lowest order
        production rates.}.
The reason to this is that at high $k_T$ the highest temperatures
dominate and they are not changed.
The higher production rate of photons in hadron gas than in
plasma \cite{Thoma} leads to a higher yield for EoS\,H
and IS\,1 than for EoS\,A and IS\,1 even though the initial temperature
is higher in the latter case (cf.~table~\ref{lampo}). The spectrum reaches
the upper limits of experimental data points around $k_T\sim 2$ GeV when
EoS\,H and IS\,1 are used.  The choice of initially cooler initial state
IS\,2 with EoS\,H decreases the yield sufficiently in $k_T\sim 2$ GeV
region but also creates a distinctive convex shape of the spectrum.

To get a better understanding of the total photon yield in different
cases we have parameterized the pQCD photon spectrum of \cite{WW} as
presented in \cite{Gale} and added it to the thermal yield.  Resulting
spectra are displayed as the upper set of curves in fig.~4 for $k_T>1.2$
GeV.  The differences are small both in normalization and shape.  We
feel that with the present accuracy of the data a $\chi^2$ fitting would
not be useful in ruling out any of the curves.  The hadronic
alternative, EoS\,H, with high temperature initial state IS\,1 tends to
stay above the data.  The hadronic EoS with the lower density initial
state, IS\,2, goes through the data but there is a tendency for the
curve to be slightly steeper than the data.  The spectrum obtained by
assuming a phase transition, EoS\,A and initial state IS\,1 gives a very
good description of data and seems, in particular in the shape, to be
closest to the data.

We should like to point out that even with the initial state IS\,2 the
maximum value of the initial temperature in the hadron gas alternative
is 245 MeV (cf.~table~\ref{lampo}).  We have checked that for the high-$k_T$
region, contributions from the short time interval with highest
temperatures dominate.  In all scenarios the contribution from matter
with temperature $T<200$ MeV is roughly 40 \% of the total thermal yield
at $k_T\simeq 3$~GeV.

\section{Discussion and conclusions}

We have studied in detail the dependence of thermal photon emission in
central lead-lead collisions at CERN SPS describing hydrodynamically the
expansion of produced matter and using known emission rates in hot
matter.  We have considered emission both from QGP and hadron gas and
compared with each other the different contributions to the thermal rate.
In the hydrodynamic calculations we have used hadronic spectra to
constrain the initial conditions including the longitudinal structure.
The significance of initial conditions is emphasized by the fact that
most of the photons, in particular at high $k_T$, are emitted in a
short time interval right in the beginning of the expansion.  For
comparison we performed also the boost invariant calculation.

We find that the measured photons can be described in terms of a thermal
model with hydrodynamical expansion either by using EoS\,A with phase
transition or EoS\,H with hadron gas phase only.  The model is
constrained to reproduce the transverse and longitudinal hadron spectra.
However, in describing the photon emission a high temperature initial
state must be assumed.  With the longitudinal scaling expansion this is
achieved by choosing a short initial time $\tau_0=0.3$ fm/c which
is not in complete accordance with longitudinal geometry.  The need for
high temperature initial state can be avoided by assuming a non-zero
initial transverse velocity in case BI\,(1.0).  We have tried this but,
e.g.~the use of a linearly increasing initial velocity as in
ref.~\cite{Peressounko} with $v_r(r=R_A)=0.3$, which produces very
closely the same photon spectrum as IS\,1 or BI(0.3) in fig.~3, leads to
a $\pi^0$ spectrum well above the case IS\,1, EoS\,H in fig.~1.  We
cannot lower the $\pi^0$ spectrum by simply decreasing the initial
temperature since we would then go below the hadron multiplicities.

The shape of the photon spectrum clearly depends on whether phase
transition is assumed or not. Unfortunately this dependence can not
be used yet to differentiate between scenarios since the differences
become as small or smaller than the experimental errors
when the initial pQCD photons are added to the spectrum. One
may argue that after the addition of initial pQCD contribution the
spectrum with EoS\,A and IS\,1 still reproduces the shape best, but the
differences are small.

With QGP formation we find the higher-loop contributions essential for
reproducing the data over the whole observed range.  In particular, the
shape of the spectrum follows the shape of the data and the
normalization leaves room for the initial pQCD photons which are
estimated to be of order 1/3 of the data with spectral shape very
similar to that of the data.  On the other hand, the experimental data
rules out, within the thermal model, an increase of the rate by a large
factor.  This could be interpreted as lending support for the assumption
that the emission rate of photons from plasma can be calculated
perturbatively already in the temperature range $200\ldots 250$ MeV and
the main part of the rate is given by the full order of
$\alpha_s$ result \cite{Arnold:hep-ph/0109064,Arnold:hep-ph/0111107}.

In summary, the WA98 photon data supports a realistic thermal model
with high density and temperature initial state constructed to describe
the main features of hadron data.
The spectra obtained by assuming a phase transition, EoS\,A, or pure
hadron gas, EoS\,H, deviate somewhat in properties but the accuracy of
the present data and the present understanding of initial pQCD photons
is not enough to rule out either alternative.

\vspace{1cm}
{\it Acknowledgements:} We are grateful to P.~Arnold, G.~D.~Moore and
L.~G.~Yaffe for sending us a parameterization of their rates,
P.~Aurenche, F.~Gelis and H.~Zaraket for explaining theirs
and M.~Thoma for discussions and comments.
This work was partly supported by the U.S. Department of Energy under
Contract No.\ DE-AC03-76SF00098 and Grant No.\ DE-FG02-87ER40328 and
by the Research Council for Natural Sciences and Engineering of
the Academy of Finland. P.H.~thanks the University of Jyv\"askyl\"a for
hospitality while writing this paper.

\end{document}